\documentclass[reprint,amsmath, amssymb, aps,]{revtex4-2}
\usepackage{amsmath,amssymb}
\usepackage{graphicx}
\usepackage{hyperref}
\usepackage{subcaption}  
\usepackage{booktabs}

\usepackage{ragged2e}
\makeatletter
\renewcommand{\fnum@figure}{\textbf{Figure~\thefigure}}
\renewcommand{\fnum@table}{\textbf{Table~\thetable}}
\renewcommand{\@makecaption}[2]{%
  \vskip\abovecaptionskip
  \sbox\@tempboxa{\textbf{#1.} #2}%
  \ifdim \wd\@tempboxa >\hsize
    \textbf{#1.} \justifying #2\par
  \else
    \global\@minipagefalse
    \hb@xt@\hsize{\hfil\box\@tempboxa\hfil}%
  \fi
  \vskip\belowcaptionskip}
\makeatother

\begin{document}
\title{Robustness Analysis and Controller Design of Arm-locking System in Space-based Gravitational Wave Detectors}

\author{Yongbin Shao}
\email{bin\_bin9267@163.com}
\affiliation{Sino-European Institute of Aviation Engineering, Civil Aviation University of China, Tianjin 300300, China}

\author{Xinyi Zhao}
\email{xy-zhao@cauc.edu.cn}
\affiliation{Sino-European Institute of Aviation Engineering, Civil Aviation University of China, Tianjin 300300, China}

\author{Long Ma}
\email{longma@cauc.edu.cn}
\affiliation{Sino-European Institute of Aviation Engineering, Civil Aviation University of China, Tianjin 300300, China}

\author{Ming Xin}
\email{xinm@tju.edu.cn}
\affiliation{School of Electrical and Information Engineering, Tianjin University, Tianjin 300300, China}
\date{\today}

\begin{abstract}
Arm-locking frequency stabilization is a key technique for suppressing laser frequency noise in space-based gravitational-wave detectors. The robustness of the arm-locking control loop is crucial for maintaining laser frequency stability, which directly impacts the accuracy of gravitational-wave measurements.
In this work, a parametric stability analysis framework is developed by combining the D-subdivision theory with the Semi-Discretization method to map the stability regions of arm-locking systems in the parameter space and identify their critical stability boundaries. Based on the frequency-domain characteristics, a robust arm-locking controller is designed to enhance loop stability under parameter perturbations. Theoretical analysis and time-domain simulations confirm that the proposed controller maintains closed-loop stability and realize suppression of laser frequency noise against parameter perturbation. 
\end{abstract}
\maketitle
\section{Introduction}
Space-based gravitational-wave observatories—such as LISA~\cite{amaro2023astrophysics, Jennrich_2009}, Taiji~\cite{luo2021taiji, hu2017taiji}, TianQin~\cite{mei2021tianqin}, and DECIGO~\cite{kawamura2021current}—employ heterodyne laser interferometry to measure picometer-level displacements over arm lengths of millions of kilometers. A major technical challenge in these missions is the suppression of laser frequency noise, which can severely limit the measurement sensitivity~\cite{ghosh2022arm}. Arm locking is a key technique that stabilizes the laser frequency by referencing the interferometer arm lengths through a closed-loop control architecture. However, the achievable performance is ultimately constrained by the stability of the arm-locking control loop itself.

Since the concept of arm locking was first introduced by Sheard et al.~\cite{sheard2003laser}, extensive studies have focused on developing the arm-locking architecture, as illustrated in Table~\ref{tab:frequency_response_Arm}. The dual arm-locking configuration mitigates the noise amplification near the null frequency inherent in the single arm-locking configuration~\cite{sutton2008laser}, while the modified dual arm-locking configuration further compensates for the Doppler-induced frequency pulling effect~\cite{mckenzie2009performance}. Recently, optical frequency-comb technology has also been incorporated to further enhance the noise-suppression performance of single arm-locking configuration~\cite{wu2022arm}. From the control-design perspective, general design procedures have been developed to satisfy both the noise suppression and phase-margin requirements of the control loop~\cite{sheard2003laser,mckenzie2009performance}. With appropriate tuning of loop gain and filter parameters, the common-arm configuration can achieve performance comparable to that of the modified dual arm-locking scheme~\cite{ke2023suppression}. More recently, beyond controller optimization, Zhang et al.~\cite{zhang2024transient}investigated the transient stability characteristics of arm-locking systems and proposed a stability criterion based on transient-response analysis, providing a new approach for the rapid identification of stable controller parameters. 
\begin{figure}[htbp]
    \centering
    \includegraphics[width=0.98\columnwidth]{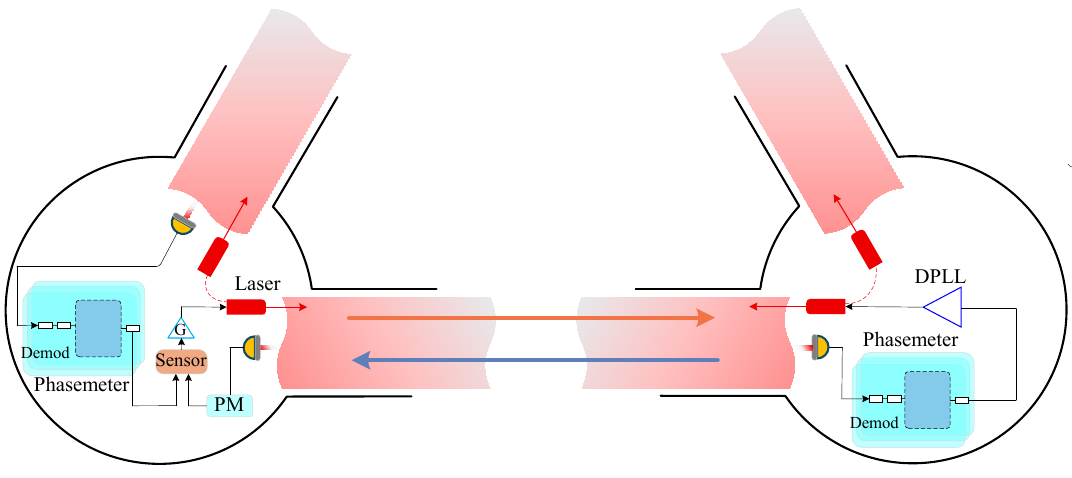}
    \caption{Conceptual diagram of the arm-locking technique. The key factor of arm-locking is the laser phase measurement, which is directly influenced by functional modules such as the phase measurement modules and the phase-locked loop (PLL) module. Other functional modules, including the telescope pointing module, can indirectly affect the laser phase measurement.}
    \label{fig:Laser_systems}
\end{figure}

However, the laser frequency stabilization system in space-based gravitational wave detectors is a complex, comprising multiple functional modules~\cite{ming2020ultraprecision}, as shown in Fig.~\ref{fig:Laser_systems}. These include the phase measurement module, the phase-locked loop (PLL), and the telescope pointing module, whose coordinated operation is essential for accurate phase measurements.
In many theoretical analyses and numerical simulations, simplified or idealized assumptions are often introduced that do not fully capture the system's realistic operational dynamics. As a result, the overall control architecture tends to have limited redundancy, making the stability of the arm-locking system highly sensitive to parameter drifts. Therefore, each functional module is crucial to be designed with adequate stability margins to tolerate parameter perturbations arising from inter-module interactions.

To address these challenges, this work establishes a parametric stability analysis framework for arm-locking systems and develops a robust controller design method under parameter perturbations. The main contributions of this study are summarized as follows: 
\begin{itemize}
    \item This work introduces a parametric stability analysis framework, applicable to the single, common, and dual arm-locking configurations, combining the D-subdivision and Semi-Discretization methods, which together provide accurate analytical boundaries and complete numerical coverage of the stability regions.
    \item The analysis reveals that arm-locking loops operate near at the stability boundary in the parameter space, accounting for their high sensitivity to multiplicative parameter perturbations. This provides a quantitative basis for evaluating robustness and guiding controller design.
    \item Building upon the above observation, we propose a robust controller design approach based on the Nyquist characteristics of the perturbed system, which compensates for stability degradation induced by parameter perturbations. The proposed framework is applied to three typical configurations single arm-locking, common arm-locking, and dual arm-locking with the dual arm-locking system serving as a representative example for controller synthesis and validation.
\end{itemize}
Sections~\ref{sec:single_arm} and ~\ref{sec:stability_region} introduce the theoretical motivation and detailed procedures of parametric stability analysis framework, while Section ~\ref{sec:Controller_Design} presents the controller design and time-domain verification results based on the dual arm-locking configuration.
\begin{table}[h]
    \centering
    \caption{Frequency response of different arm-locking configurations~\cite{mckenzie2009performance}}
    \label{tab:frequency_response_Arm}
    \begin{tabular}{ll}
    \hline
    Configuration & Frequency response \\
    \hline
    Single          & $P_s(\omega)=2i\sin(\tau_{ij}\omega) e^{-i\omega\tau_{ij}}$ \\
    Common          & $P_+(\omega)=2(1-\cos(\Delta\tau\omega)e^{-i\omega\bar{\tau}})$  \\
    Difference      & $P_-(\omega)=-2i\sin(\Delta\tau\omega)e^{-i\omega\bar{\tau}}$ \\
    Dual            & $P_d(\omega)=P_+(\omega)-\frac{p}{1+p}\frac{1}{i\omega\Delta\tau}P_-(\omega)$ \\
    Modified dual   & $P_M(\omega)=P_+(\omega)H_+(\omega)-P_-(\omega)H_-(\omega)$ \\
    \hline
    \end{tabular}
\end{table}
\section{Case Study Single-Arm Locking System}
\label{sec:single_arm}
\begin{figure*}[htb]
    \centering
    \includegraphics[width=0.9\linewidth]{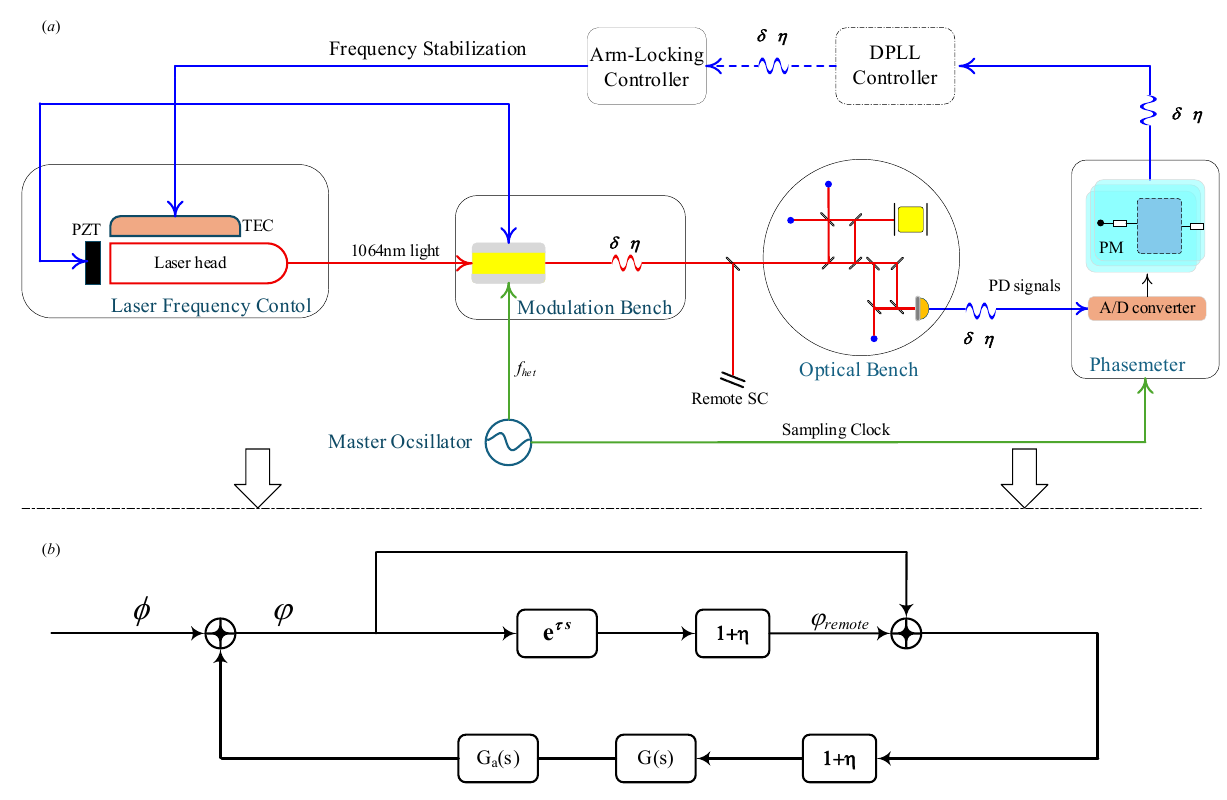}
    \caption{Arm-locking technology (a) Conceptual diagram of the arm-locking technique. 
    The arm-locking system involves several functional modules, including Laser frequency control, Optical Bench, and Phasemeter. 
    The red lines represent the optical phase of the laser beams, while the blue lines denote digital control signals. 
    Symbols $\delta$ and $\eta$ indicate potential multiplicative and additive disturbances, respectively. 
    (b) Control block diagram of the single arm-locking system. 
    Considering multiplicative disturbances, the single arm-locking control diagram includes the laser frequency actuator transfer function $G_a$ and the arm-locking controller $G(s)$.
    }
    \label{fig:subsystem}
\end{figure*}
In the implementation of arm-locking, the interactions between several functional modules involve the transmission of signals and specification of design parameters. Fig.~\ref{fig:subsystem} illustrates the modules involved in the laser arm-locking frequency stabilization process, as well as the potential points where disturbances may occur. Frequency stabilization depends on the phase signal produced by the phase measurement module, together with the weak-light phase-locking process executed by the remote spacecraft's laser phase-lock loop. Because these modules have complex internal dynamics, simplified models are generally adopted when analyzing the arm-locking scheme. However, such simplifications can produce losses of stability under certain combinations of parameter values. To assess the robustness of the arm-locking system, we introduce perturbations at the inter-module signals in the form $(1+\eta(t))X(t)+n(t)$ where $\eta(t)$ denotes multiplicative perturbations (modelling fluctuations in effective gain) and $n(t)$ captures additive perturbations (representing measurement noise), and $X(t)$ is the inter-module signal. The additive perturbations arising from the phase measurement process are interpreted as injected technical noise into the arm-locking loop, whereas the multiplicative ones correspond to gain drifts of the arm-locking controller. 
\begin{figure}[htbp]
    \centering
    \includegraphics[width=0.95\linewidth]{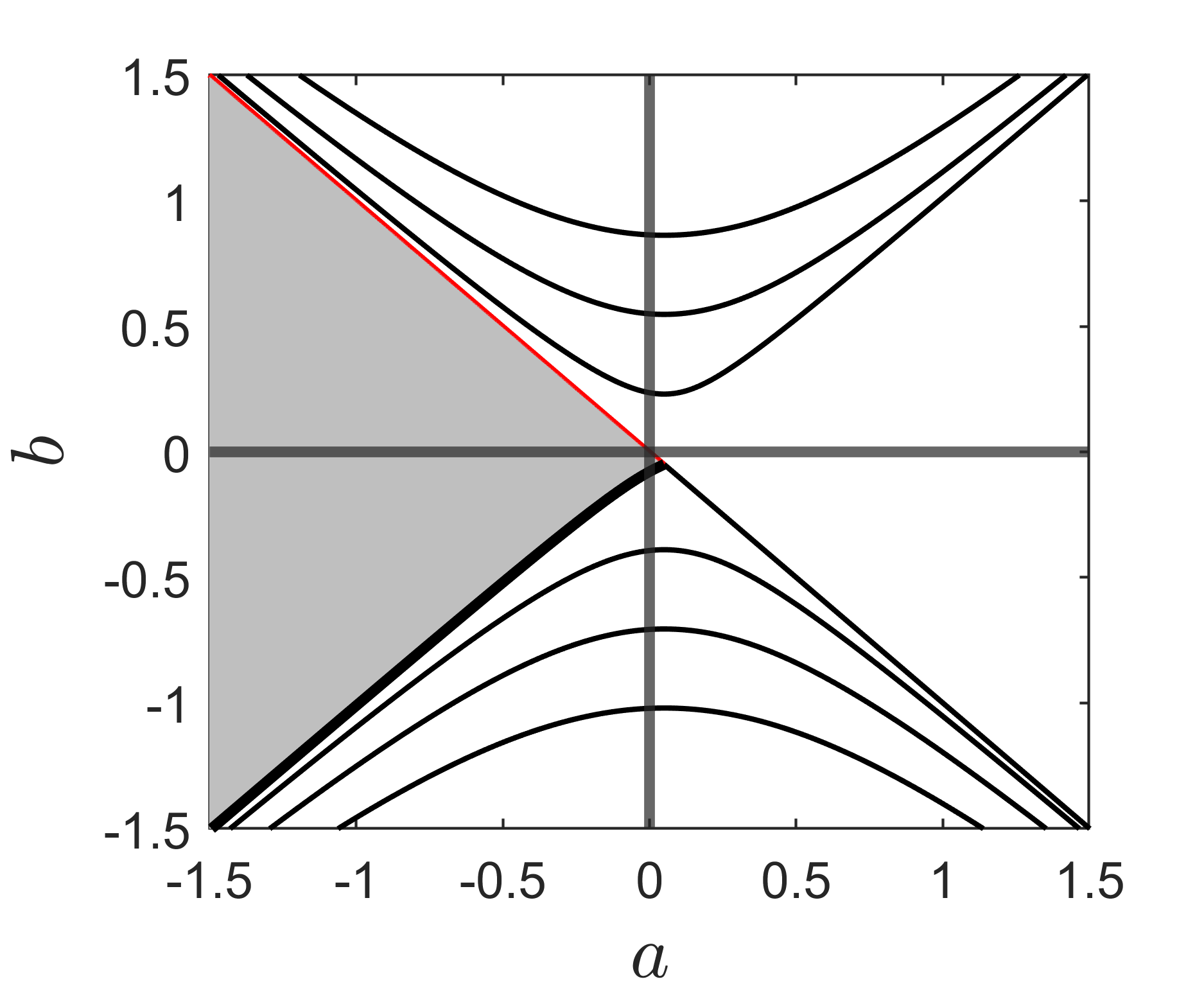}
    \caption{Stability chart of the Hayes equation. the red line (slope -1 through the origin) marks the ideal phase-locking condition $(i.e.,k2=1)$, corresponding to the single arm-locking system.}
    \label{fig:Hayes_StaChart}
\end{figure}

In this work, we consider the case that perturbations arise in the weak-light phase-locking process of the remote spacecraft. Therefore, the returned phase signal, $\varphi_{\rm remote}(t)$, can be written as
\begin{equation}
    \label{eq:fov1}
    (1+\eta(t))\varphi_{\mathrm{remote}}(t) + n(t)
\end{equation}
For the single arm-locking configuration shown in Fig.~\ref{fig:subsystem},  the time-domain equation describing how remote spacecraft parameter perturbations($\delta,\eta$) affect the laser phase noise can be written as
\begin{align}
    \label{eq:fov2}
    \varphi(t) 
    &+ k_1 \int_{-\infty}^{+\infty}
    h_c(\theta) h_a(\theta) 
    \Big(\varphi(t-\theta) \nonumber \\
    &\qquad - k_2 \varphi(t-\tau-\theta)\Big) d\theta 
    = \phi(t).
\end{align}
where $h_c(t)$ is the impulse response of the controller G, and $h_a(t)$ is the impulse response of the laser frequency actuator. To incorporate the effect of parameter perturbations, the coefficients are defined as $k_1 = 1 +n $ and $k_2 = 1 + \eta$, where $n$ and $\eta$ quantify the deviations from the nominal values.

In practice, laser frequency actuators can be implemented in several forms, such as temperature tuning, piezoelectric transducers (PZT), or electro-optic modulator (EOM). For simplicity, the actuator is modeled as an integrator, $1/s$, whose impulse response is the unit step function $u(t)$. Since this work focuses on the intrinsic structure of the arm-locking system, the use of a linear controller does not alter its essential characteristics. Therefore, a simple gain controller is adopted to facilitate the analysis of stability in the parameter space.

By differentiating Eq. (2), the convolution integral is eliminated, leading to a simplified delay-differential equation:
\begin{equation}
    \label{eq:fov3}
    \dot\varphi(t)+k_1 \left(\varphi(t)-k_2\varphi(t-\tau)\right)=0
\end{equation}
This equation has the same structure as the well-known Hayes delay-differential equation,$\dot{x}(t) = a x(t) + b x(t - \tau)$~\cite{breda2014stability}, whose stability chart is illustrated in Fig.~\ref{fig:Hayes_StaChart}, where the gray-shaded region indicates the stable domain. By comparison, we have $a = -k_1$, $b = k_1 k_2$, and thus $k_2 = -b/a $.
When $k_2 = 1$, the system corresponds to the ideal phase-locking condition without any perturbations, represented in Fig.~\ref{fig:Hayes_StaChart} by the red line of slope $-1$ through the origin. In this case, the system lies exactly on the stability boundary, and variations in $k_1$ do not affect stability directly. If $k_2 > 1$, the corresponding state points deviate from the gray region, and the system becomes unstable. By contrast, if $k_2 < 1$, the line rotates counterclockwise to the gray region, and the system remains stable. The results demonstrate that stability is determined mainly by multiplicative perturbations. Hence, in the analysis of common arm-locking and dual arm-locking  (Section III), we focus on discuss the multiplicative perturbations.

Remark: Note that Eq.~(\ref{eq:fov3}) is obtained by differentiating Eq.~(\ref{eq:fov2}), and therefore the two formulations are not strictly equivalent. The differentiation step introduces an additional root at $s=0$. This root is marginally stable, which does not affect the stability of the system.  However, this extra root changes the interpretation of the stability boundary. Consequently, the red line in Fig.~\ref{fig:Hayes_StaChart} should be understood as an analytical stability boundary, rather than as evidence of a purely imaginary eigenvalue on the axis.

\section{Stability Region Determination via Semi-Discretization and D-Subdivision Theory}
\label{sec:stability_region}
\begin{figure*}[htb]
    \centering
    \begin{subfigure}{0.3\textwidth}
        \centering
        \includegraphics[width=\linewidth]{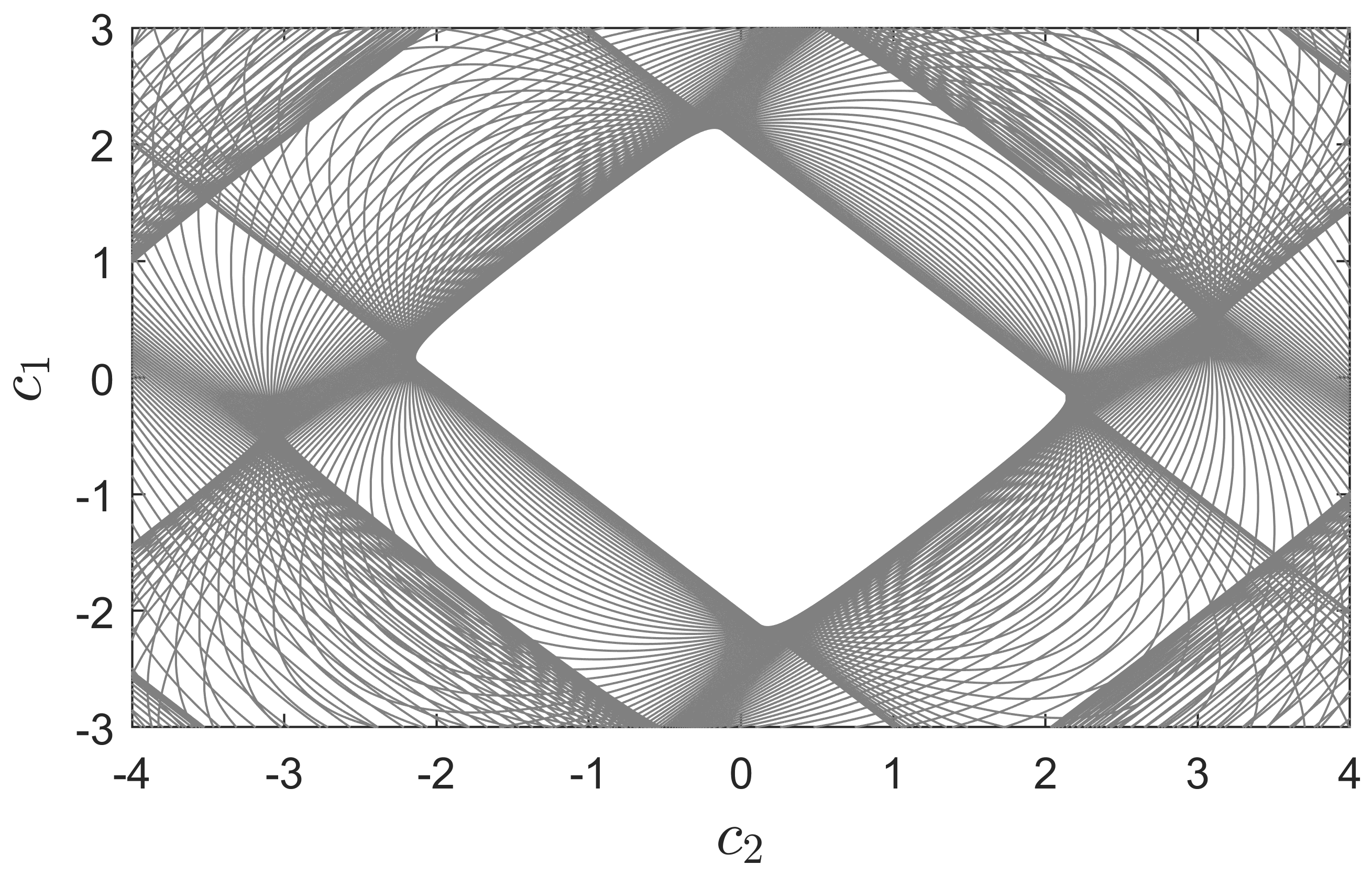}
        \caption{}
        \label{fig:Common_sub_a}
    \end{subfigure}
    \hspace{0.03\textwidth}
    \begin{subfigure}{0.3\textwidth}
        \centering
        \includegraphics[width=\linewidth]{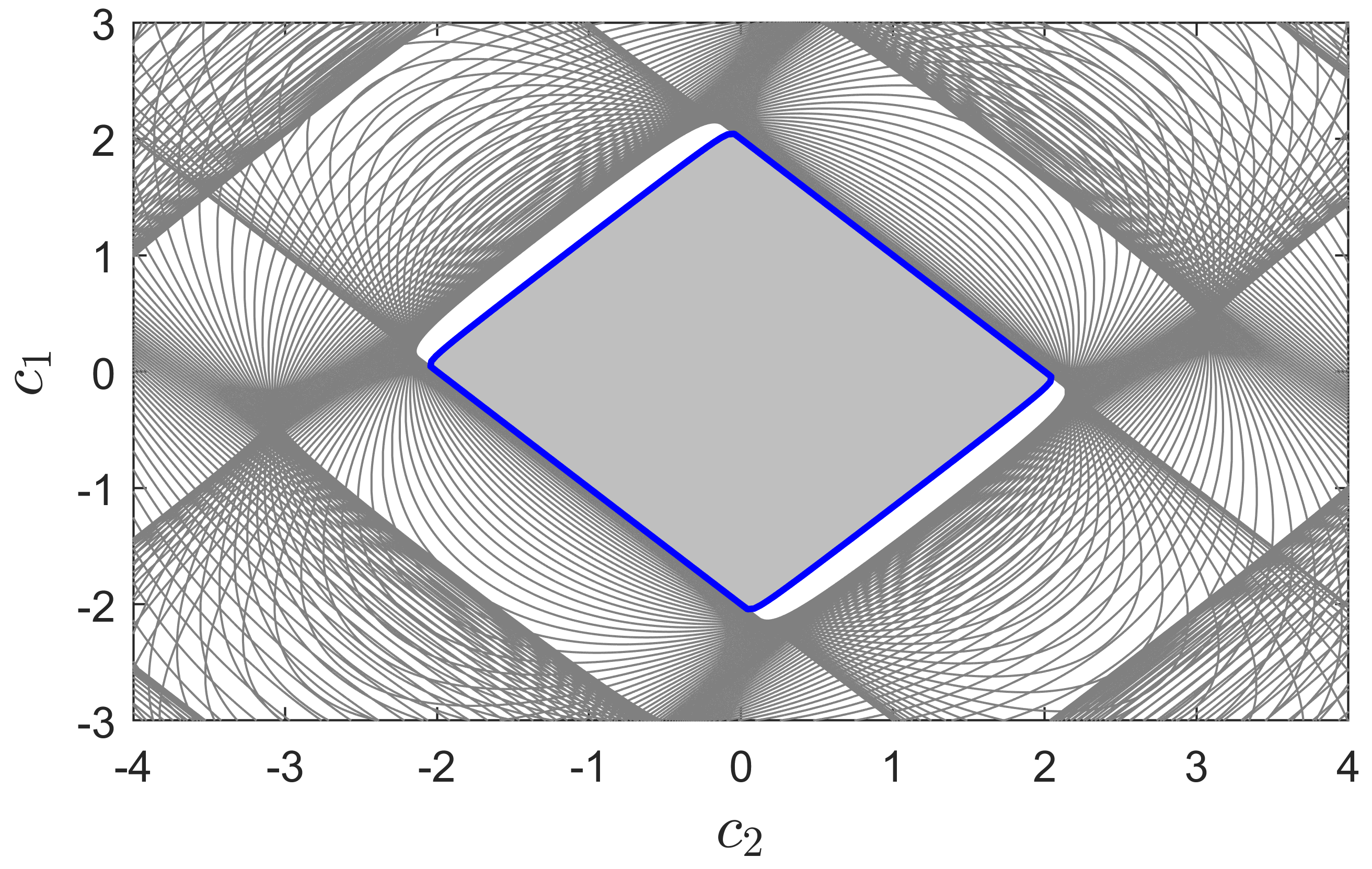}
        \caption{}
        \label{fig:Common_sub_b}
    \end{subfigure}
    \hspace{0.03\textwidth}
    \begin{subfigure}{0.3\textwidth}
        \centering
        \includegraphics[width=\linewidth]{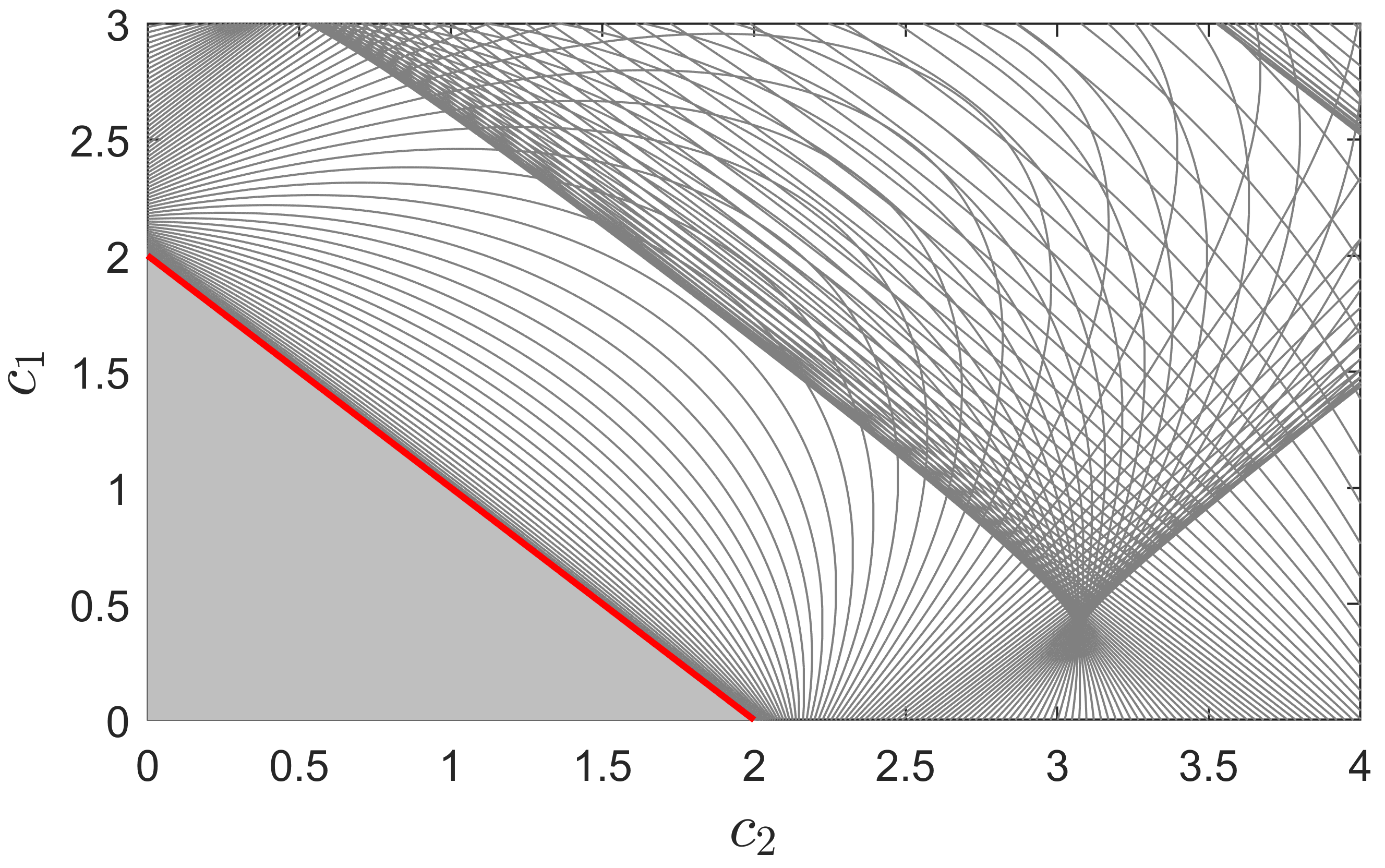}
        \caption{}
        \label{fig:Common_sub_c}
    \end{subfigure}
    \caption{
        Common arm system: (a) $c_1$-$c_2$ plane D-curves: the gray curves represent the D-curves plotted from the corresponding equations. 
        The regions separated by these curves have different numbers of unstable roots (roots with positive real parts). 
        (b) Stability region obtained using the Semi-Discretization method: 
        the dark gray area indicates the stable region computed by the Semi-Discretization method, 
        and the blue curve represents the corresponding stable boundary. The step size $h=0.05$.  
        (c) Local stability boundary of the common-mode arm system: the red straight line is given by $c_1 + c_2 = 2$, 
        and the dark gray shaded area denotes the stable region.
    }
    \label{fig:Disc_Sub}
\end{figure*}
\subsection{Common Arm}
\label{sec:common_arm}
In this section, we extend the analysis to the common arm-locking configuration. Specifically, the loop is described by an overall controller gain $g$ and $\tau_1<\tau_2$ denote the round-trip delays of the two arms. Moreover, we introduce two coefficients $(c_1,c_2)$ that represent multiplicative perturbations associated with the respective arms. Note that such perturbations do not necessarily originate from the loop itself, but for convenience they are treated as perturbations of the overall scheme.
Based on this, the open-loop transfer function can then be written as
\begin{equation}
    \label{eq:common_arm_tf}
    P_+(s) = \frac{g}{s}\left(2-c_1e^{-\tau_1 s}-c_2e^{-\tau_2 s}\right) 
\end{equation}
Applying the D-subdivision method~\cite{breda2014stability} with $s = j\omega$ yields the following parametric equations of the stability boundary
\begin{subequations} 
    \label{eq:5a}
    \begin{align}
        \begin{cases}
            c_1(\omega)= \frac{2\sin(\omega\tau_2)+\frac{\omega}{g}\cos(\omega\tau_2)}{\sin(\omega(\tau_2-\tau_1))} \\
            c_2(\omega)= \frac{-2\sin(\omega\tau_1)-\frac{\omega}{g}\cos(\omega\tau_1)}{\sin(\omega(\tau_2-\tau_1))}
        \end{cases}, 
        \quad \omega \neq 0 
    \end{align}
\end{subequations}
with the special case
\begin{equation}
    \label{eq:5b}
    c_1 + c_2 = 2, \quad \omega = 0 \tag{5b}
\end{equation}

In the single arm-locking configuration, this process directly yields the stability region together with its analytical boundaries(see Eq.~(\ref{eq:fov2}) and Fig.~\ref{fig:Hayes_StaChart}). However, in dual arm-locking or common arm-locking, the presence of multiple delays leads the stability boundaries (D-curves) exhibiting a more complex geometry, including multiple branches, cusps, and ring-like structures. As an illustrative example, the D-curve is computed using $\tau_1 = 20.1$, $\tau_2 = 19.9$, and $g = 1$, and the resulting curve is shown in Fig.~\ref{fig:Common_sub_a}. The selection of the delay $\tau$ is determined by the arm length of the Taiji interferometer.~\cite{liu2021numerical}. 

Although D-subdivision theory yields closed-form analytical stability boundaries, its practical application suffers from numerical implementation, such as the sampling, numerical precision, finite domain, which may fail to accurately obtain the complete stability boundary. In addition, the determination of the stable region is also sensitive to the accuracy of the numerical solution. 

To address these challenges, this work adopts the Semi-Discretization method~\cite{insperger2011semi} as a complementary numerical tool for identifying stability regions. Compared with standard discretization methods for time-delay systems, Semi-Discretization offers higher accuracy and a more intuitive computational procedure. However, its representation of the stability boundary is often less precise. By combining Semi-Discretization with D-subdivision theory, the advantages of both approaches are leveraged: D-subdivision captures the geometric features of the stability boundaries, while Semi-Discretization guarantees that the entire stability region is covered. The proposed method yields a stability analysis that is both more accurate and more comprehensive.

Recalling Eq.~(\ref{eq:common_arm_tf}), the corresponding homogeneous delay differential equation:
\begin{equation}
    \label{eq:common_arm_DDE}
    \dot{\varphi}(t)+2g\varphi-gc_1\varphi(t-\tau_1)-gc_2\varphi(t-\tau_2)=0
\end{equation}

To facilitate further analysis, we apply the Semi-Discretization method. In this approach, the derivative term remains continuous, while the delayed terms $\varphi(t-\tau_1)$ and $\varphi(t-\tau_2)$ are discretized.
With a step size $h \ll \tau_k, (k=1,2)$ and discrete time instants $t_i = i h$, the delays are written as $ \tau_k = r_k h$, yielding the recurrence relation
\begin{equation}
    \label{eq:common_arm_Disc}
    \varphi(t_{i+1}) = P \, \varphi_{t_i} + R \, \varphi(t_{i-r_1}) + g h c_2 \, \varphi(t_{i-r_2})
\end{equation}
where
\begin{equation}
    P = e^{-2 g h}, \quad 
    R = \int_{t_i}^{t_{i+1}} g c_k e^{-2 g (t_i - \nu)} \, d\nu \notag
\end{equation}
For convenience, we denote $\varphi(t_i) = \varphi_i$. For $i>r_2$, Eq.~(\ref{eq:common_arm_Disc}) extends to an augmented state-pace form:
\[
\label{eq:Mstate}
\begin{bmatrix}

    \varphi_{i+1} \\
    \varphi_i \\
    \vdots \\
    \varphi_{i - r_1 + 1} \\
    \vdots \\
    \varphi_{i - r_2 + 1}
\end{bmatrix}
    =
    \underbrace{
        \begin{bmatrix}
            P & 0 & \cdots & R_1 & \cdots & R_2 \\
            I & 0 & 0 & \cdots & 0 & 0 \\
            0 & I & 0 & \cdots & 0 & 0 \\
            \vdots &\vdots & \ddots & & \vdots & \vdots \\
            \vdots &\vdots & & \ddots & \vdots & \vdots \\
            0 & 0& \cdots & 0 & I & 0
        \end{bmatrix}
    }_{G_\Phi}
    \begin{bmatrix}
        \varphi_i \\
        \varphi_{i - 1} \\
        \vdots \\
        \varphi_{i - r_1} \\
        \vdots \\
        \varphi_{i - r_2}
    \end{bmatrix}
    \tag{8}
\]
where $G_\Phi$ is the corresponding state-transition matrix constructed from $P, R_1, R_2$. It's obvious that the system (Eq.~\ref{eq:Mstate}) is stable if and only if the spectral radius of the corresponding state-transition matrix is less than unity. 

In Fig.~\ref{fig:Common_sub_b}, the gray region denotes the stability domain obtained using Semi-Discretization, while the blue curve marks its stability boundary. For comparison, the D-subdivision analysis yields the analytical stability boundary of the common arm-locking as the straight line $c_1 + c_2 = 2$, shown in Fig.~\ref{fig:Common_sub_c} in red. We can observe that under the ideal condition $c_1 = c_2 = 1$, the common arm-locking system lies exactly on this stability boundary.

\begin{figure}[htbp]
    \centering
    \includegraphics[width=0.8\linewidth]{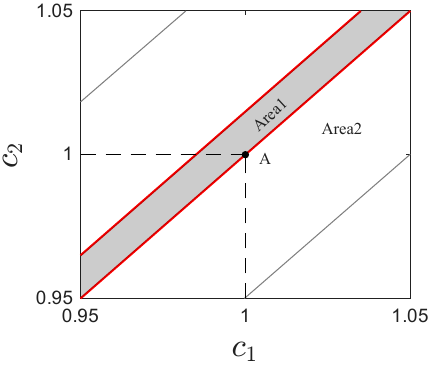}
    \caption{Stability chart of the dual arm locking system when $g  = 10$. The gray-shaded region (Area1) represents the stable domain, 
    while the Point A(1,1) indicates the location corresponding to the ideal dual arm-locking system.
    The red line represents the stability boundary obtained from the proposed method.}
    \label{fig:dual_arm_StaChart}
\end{figure}
\subsection{Dual Arm}
Compared with the common and single arm-locking systems, the dual arm-locking system is more complex, as it forms its feedback signal as a frequency-dependent linear combination of the two delayed arm measurements. Consistent with our notation, $c_1$ and $c_2$ are introduced as multiplicative perturbations on each arm (nominally $c_1=c_2=1$).
Its open-loop transfer function can be expressed as follows:
\begin{align}
P_d(j\omega) &=\frac{g}{j\omega}(2-c_1e^{-\tau_1j\omega}-c_2e^{-\tau_2j\omega}) \notag \\ 
&+\frac{gp}{\omega+p}\frac{1}{\omega^2\Delta\tau}(c_1e^{-\tau_1\omega}-c_2e^{-\tau_2\omega})
\end{align}
where $\Delta\tau = (\tau_1 - \tau_2)$, $g$ and $p$ denote the overall gain and the corner frequency of the controller, respectively. Following the methodology established in Sec.~\ref{sec:common_arm}, the combined D-subdivision and Semi-Discretization approach is applied to the dual arm-locking configuration, providing a consistent and comprehensive framework for stability analysis in dual arm-locking system. 
Using representative parameters $\tau_1=20.1,\tau_2=19.9 $ and $g=10$, the resulting stability diagram in the $(c_1,c_2)$ plane is shown in Fig.~\ref{fig:dual_arm_StaChart}.

\begin{figure}[htbp]
    \centering
    \includegraphics[width=0.8\linewidth]{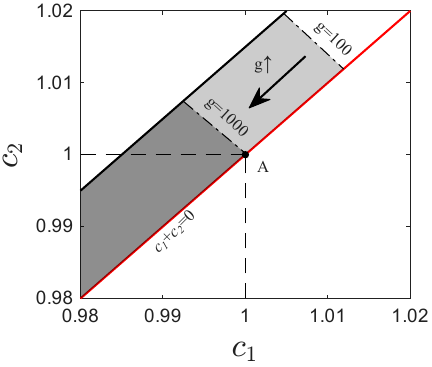}
    \caption{Stability chart of the dual arm-locking system as the gain $g$ increases. The gray-shaded area indicates the stable domain. As $g$ increases, the stability boundary represented by the blue dashed line gradually shrinks, while the boundary denoted by the red line remains unchanged (note that the upper boundary of the stable region, black solid line, is not the exact stability limit; precise computation is not performed since the arm-locking parameters never reach this boundary).}
    \label{fig:dual_arm_stability2}
\end{figure}

As the gain increases, the stability region gradually shrinks as Fig~\ref{fig:dual_arm_stability2} shows. 
This is because perturbations of the parameters $c_1$ and $c_2$ may cause the Nyquist curve of the dual arm-locking system to cross the negative real axis in the complex plane. 
Consequently, the system does not possess infinite gain margin. 
Therefore, as the gain increases, the upper-right part of the shaded region in the figure becomes smaller.
\section{Robustness Analysis and Controller Design}
\label{sec:Controller_Design}
\begin{figure*}[htb]
    \centering
    \begin{subfigure}{0.24\textwidth}
        \centering
        \includegraphics[width=\linewidth]{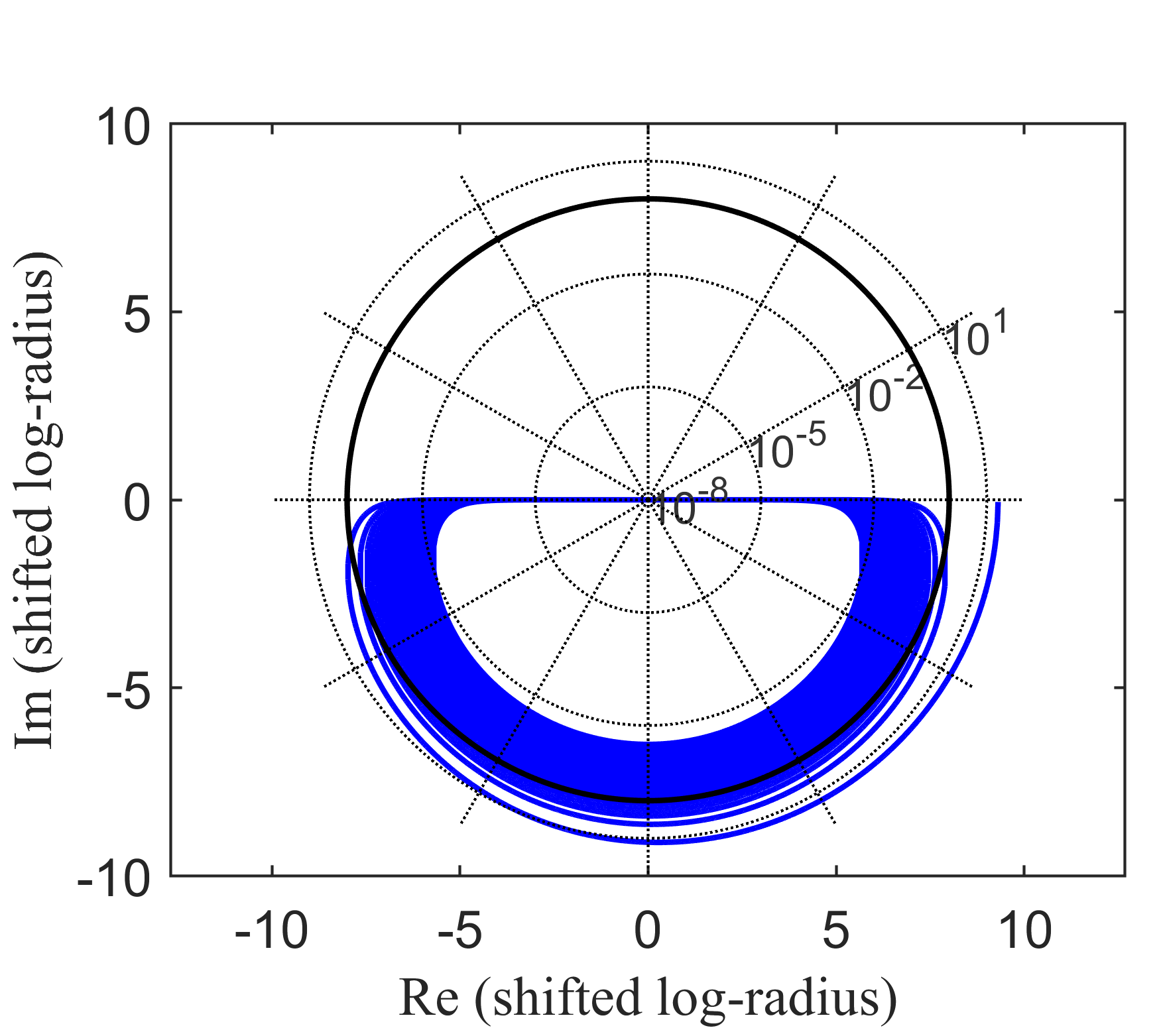}
        \caption{}
        \label{fig:sub_a}
    \end{subfigure}
    \hfill
    \begin{subfigure}{0.24\textwidth}
        \centering
        \includegraphics[width=\linewidth]{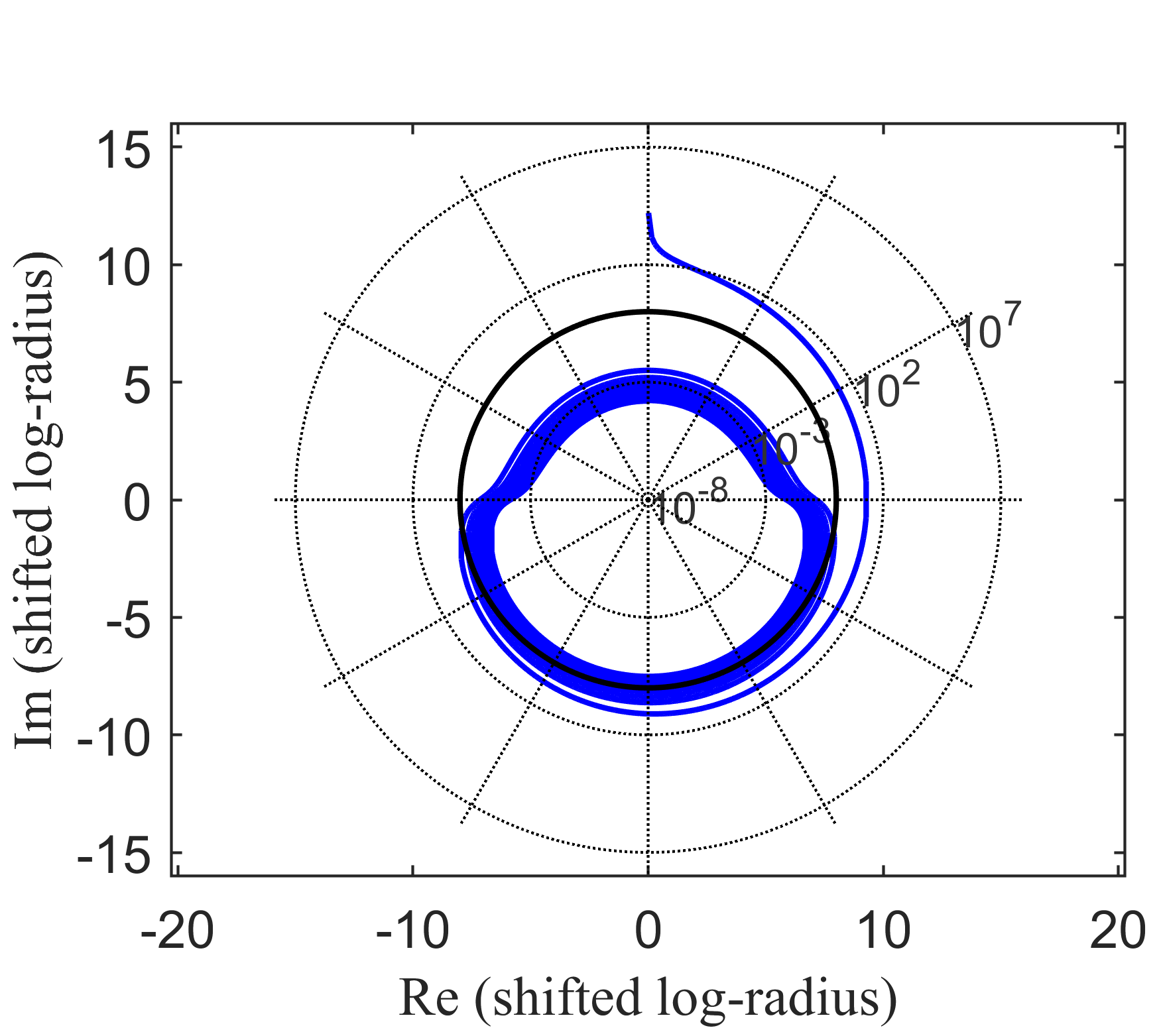}
        \caption{}
        \label{fig:sub_b}
    \end{subfigure}
    \hfill
    \begin{subfigure}{0.24\textwidth}
        \centering
        \includegraphics[width=\linewidth]{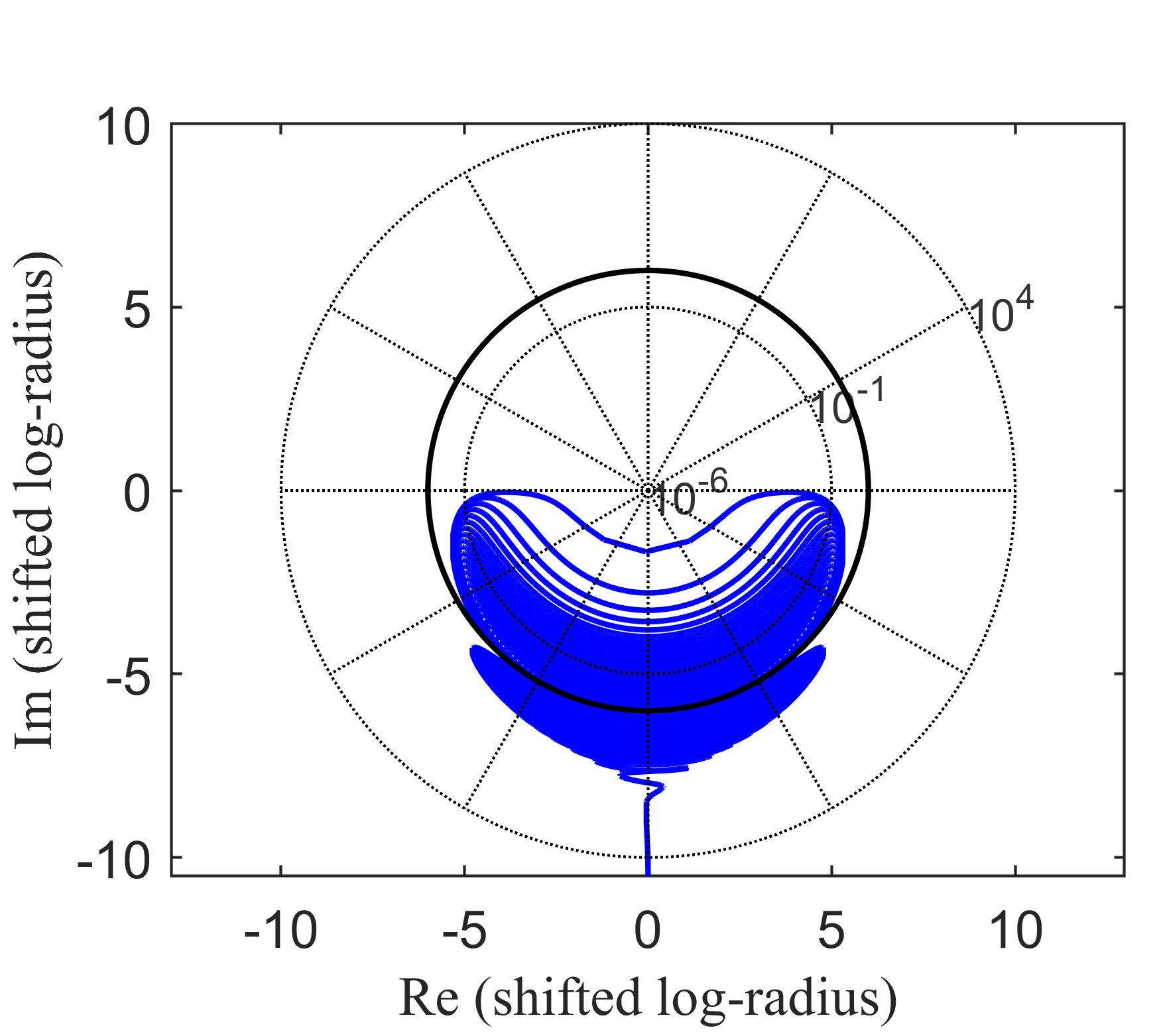}
        \caption{}
        \label{fig:sub_c}
    \end{subfigure}
    \hfill
    \begin{subfigure}{0.24\textwidth}
        \centering
        \includegraphics[width=\linewidth]{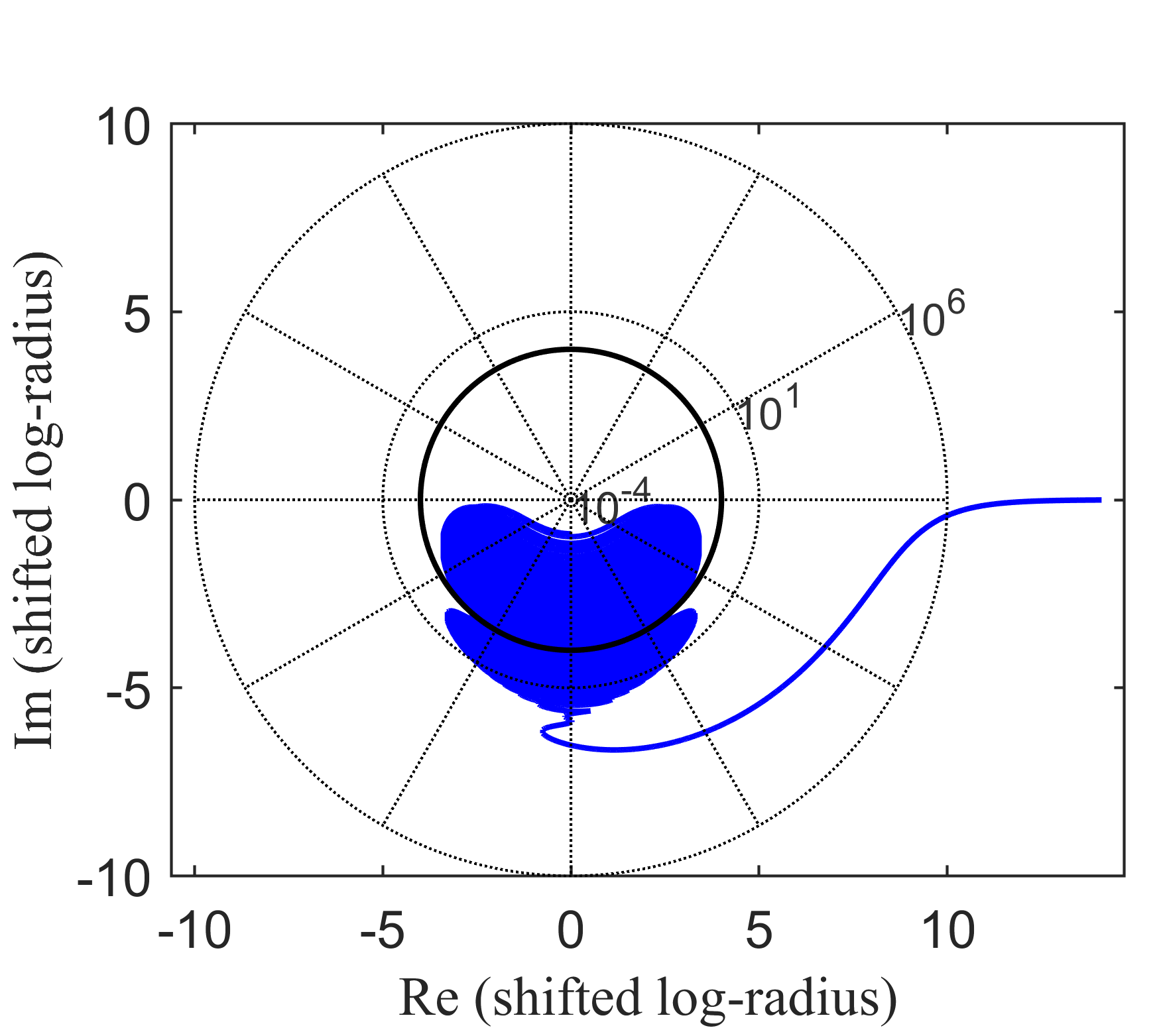}
        \caption{}
        \label{fig:sub_d}
    \end{subfigure}
         \captionsetup{
        justification=raggedright, 
        singlelinecheck=false       
    }
    \caption{Nyquist plots of the arm-locking system: (a) Nyquist curve of the single arm-locking system with $c_1 = 1$
    (b) Nyquist curve of the single arm-locking system with $c_1 = 1.001$. 
    (c) Nyquist curve of the dual arm-locking system at the stability boundary with $(c_1, c_2) = (1, 1)$. 
    (d) Nyquist curve of the dual arm-locking system in Area~2 with $(c_1, c_2) = (1, 0.995)$.}
    \label{fig:nyquist_four}
\end{figure*}
\subsection{Frequency Domain Analysis}
Building on the time-domain stability analysis presented in Sec.~\ref{sec:single_arm}, this section analyzes the frequency-domain characteristics of the single, common, and dual arm-locking systems under parameter perturbations. The frequency response offers additional physical insight that guides the design of more robust controllers discussed later in this paper. In practice, perturbations in $c_1$ and $c_1$ can arise from various sources. In this work, their magnitude is limited to $10^{-6}$, with the rationale for this choice and its physical origins discussed in the Appendix~\ref{sec:appendix}. 

For the single arm-locking configuration, the open-loop transfer function is $P_s(j\omega) = g(1 - c_1 e^{-\tau j\omega})/s$. When $c_1 = 1$ (ideal phase locking), the Nyquist curve plotted is tangent to the origin in Fig.~\ref{fig:sub_a}, indicating an infinite gain margin. In the presence of small multiplicative perturbations, $c_1$ may slightly exceed $1$. Under this condition, the frequency-domain property of the single arm-locking system is altered: an infinite branch appears at $\omega = 0$ as illustrated by the Nyquist plot in Fig.~\ref{fig:sub_b}. As the frequency varies from $\omega = 0^-$ to $\omega = 0^+$, the phase of the Nyquist curve changes from $-\pi/2$ to $\pi/2$, and the infinite branch encloses the entire left half of the complex plane. Moreover, when $c_1 > 1$, we have $P_s(2\pi/\tau) = 1 - c_1 < 0$. The Nyquist curve no longer crosses the real axis at the origin; instead, the crossing has moved to the negative real axis in the complex plane.
So the gain margin is no longer infinite and becomes dependent on the magnitude of the perturbation. 

For the common arm-locking system, the Nyquist curve is also tangent to the origin in the complex plane. Consequently, the frequency-domain behavior leading to instability is similar to that of the single arm-locking system: an infinite branch at $\omega = 0$ causes instability, and the Nyquist curve crossing the real axis results in a reduced gain margin.
 
For the dual arm-locking configuration, the same stability characteristics are preserved under ideal conditions-i.e., the system operates precisely on the analytical stability boundary shown in Fig.~\ref{fig:dual_arm_StaChart}. When multiplicative perturbations are introduced in the two arms, the operating point $(c1,c2)$ deviates from the boundary (e.g., to $(1,0.995)$ in Region II of in Fig.~\ref{fig:dual_arm_StaChart}), resulting in loss of stability. The corresponding Nyquist plot is shown Fig.~\ref{fig:sub_d}. According to the Nyquist stability criterion, in the neighborhood of $s = 0$, the open-loop transfer function $P_d(s)\sim O(1/s^2)$. Since $P_d$ has a pole at $s = 0$, letting $s = r e^{j\theta}$ and taking the limit $r \to 0$, the corresponding Nyquist trajectory develops an infinite branch that encircles the origin once in the counterclockwise direction as $\omega$ passes through zero, indicating instability. Moreover, the dual arm-locking loop exhibits only a finite gain margin, which decreases with increasing perturbation amplitude. Therefore, even small deviations in $c_1$ and $c_2$ may significantly alter the stability margins, careful consideration of these effects is required in controller design.
\begin{figure}[htb]
    \centering
    \begin{subfigure}{0.47\linewidth}
        \centering
        \includegraphics[width=\linewidth]{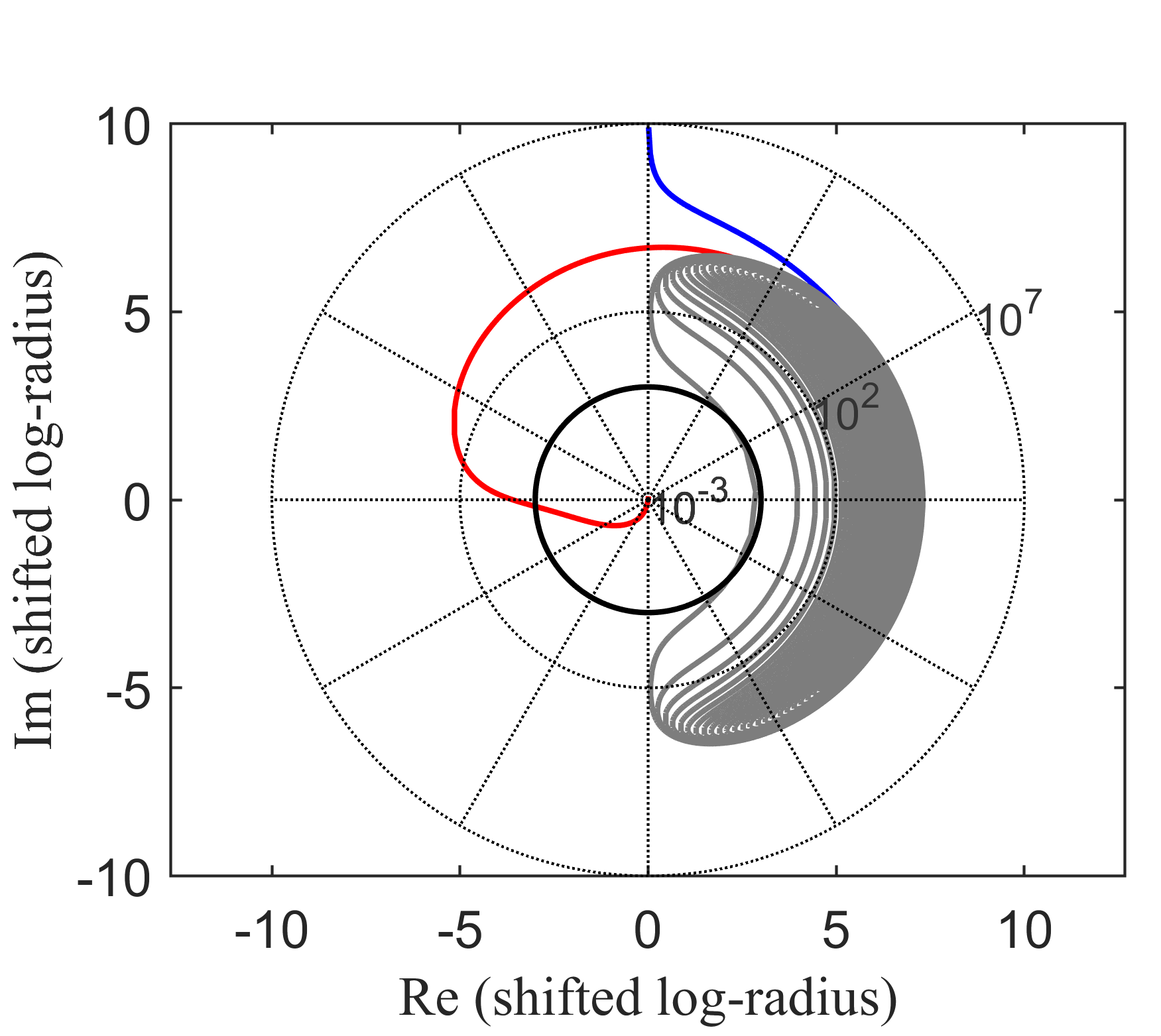}
        \caption{}
        \label{fig:col_a}
    \end{subfigure}
   \hfill
    \begin{subfigure}{0.47\linewidth}
        \centering
        \includegraphics[width=\linewidth]{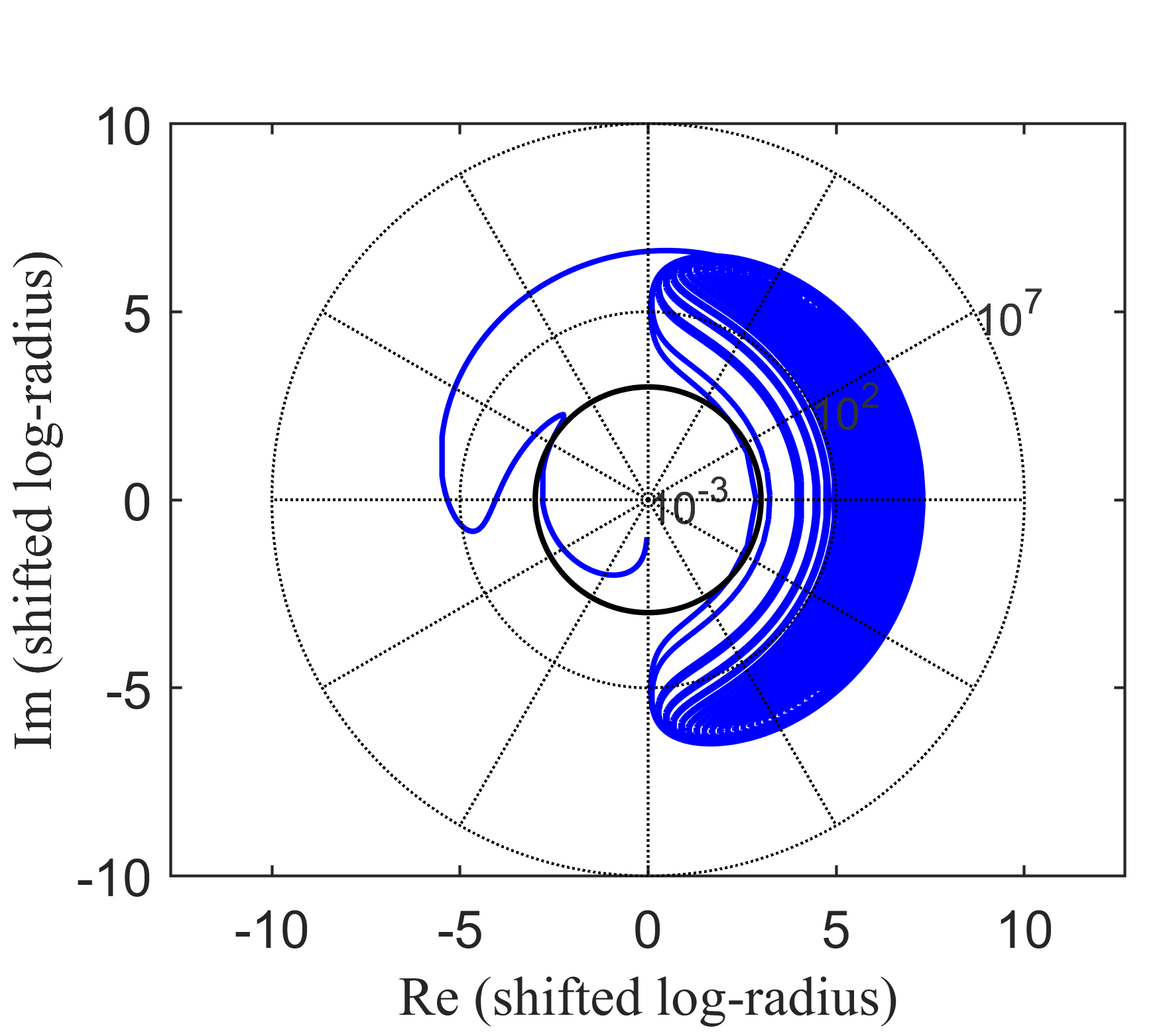}
        \caption{}
        \label{fig:col_b}
    \end{subfigure}
    \caption{Nyquist plots analysis: (a) Nyquist plot of dual arm-locking system with high-pass filter $G_H(s)$. the red curve is the Nyquist plot of the system with $G_L(s)$,
    the blue curve is the Nyquist plot of the system $P_d(s)$,  while the dark blue line indicates the overlapping high-frequency response. 
    (b) Nyquist plot of dual arm-locking system with controller $C(s)$.}
    \label{fig:two_in_onecol}
\end{figure}
\subsection{Controller Design}
\label{sec:controller_design}
Although the robustness of the arm-locking system also depends on the parameters of other functional modules, this section focuses on arm-locking controller design aspect, using the more complex dual arm-locking configuration as a representative case, to enhance system robustness based on the above frequency-domain analysis.

The gain $g$ is set to $5000$ to meet the basic noise suppression requirements. To avoid interference with the scientific measurement band ($1$ mHz - $1$ Hz), the robustness-enhancing controller is designed to act primarily below $10^{-4}$ Hz. The complete controller consists of two stages.

Stage I: To eliminate the infinite branch at low frequencies, two high-pass filters $G_L$ with poles cascaded below $0.1$mHz are introduced. However, the system remains unstable at this time. Near zero frequency, the open-loop system $P_d$ behaves approximated as $O(-s)$, and the $\pi$ phase lead contributed by the high-pass filters cause $P_d$ to encircle $(-1, 0)$ clockwise on the Nyquist plot, illustrated in Fig.~\ref{fig:col_a}.

Stage II: Phase lead compensators $G_c$ are utilized to ensure that the phase at the unity gain frequency $f_{ug}$ remains below $\pi$. Two additional crossover frequencies $f_{T,1}$, $f_{T,2}$ located on the negative real axis, are introduced to prevent the Nyquist trajectory from enclosing the $(-1, 0)$ point. The resulting stable configuration is represented by the blue curve in Fig.~\ref{fig:col_b}. 

\begin{figure}[h]
    \centering
    \includegraphics[width=\linewidth]{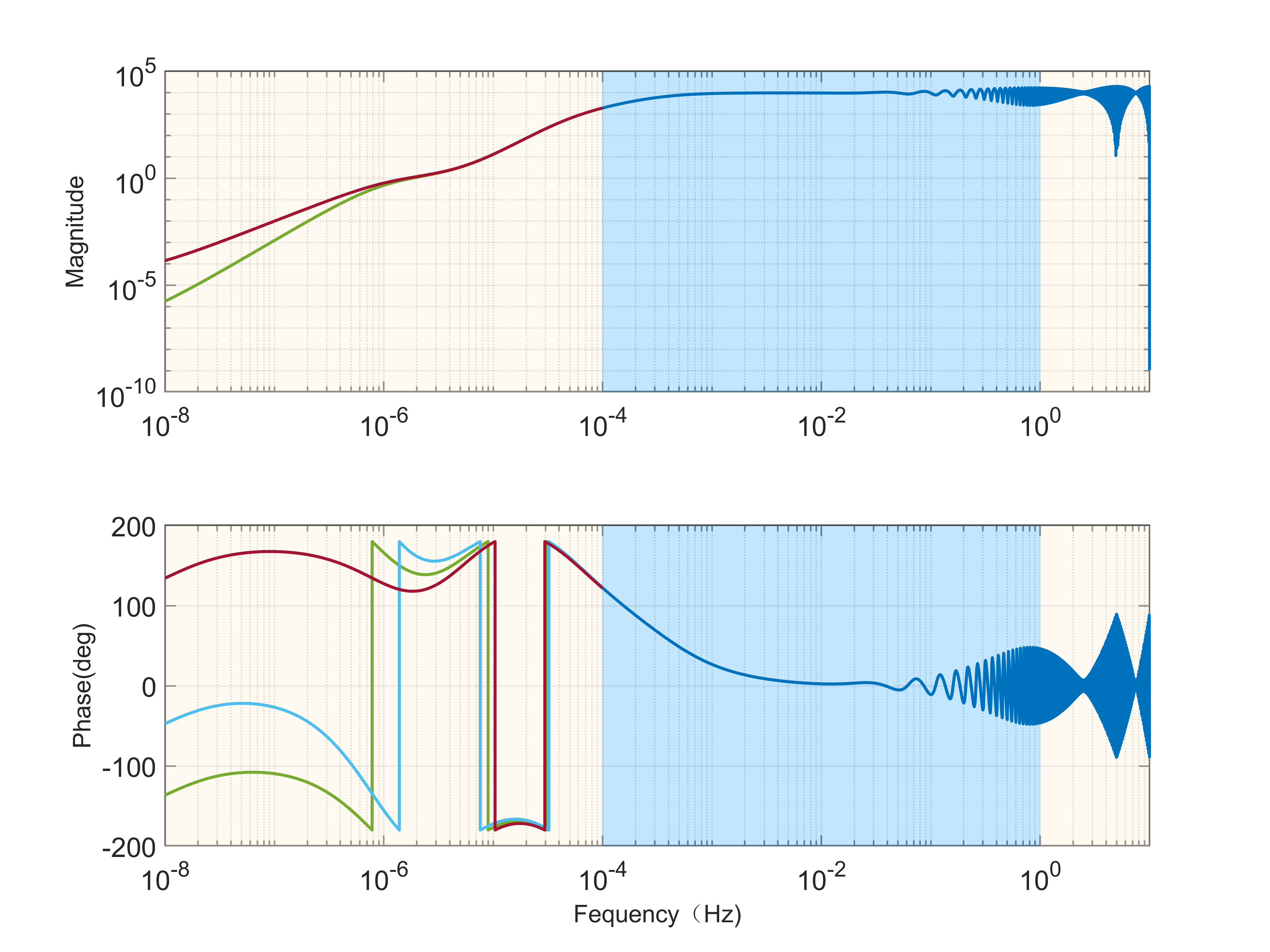}
    \caption{Bode plot of the dual arm-locking locking system with controller $C(s)$ under parameter perturbations. 
    The red line represents the perturbed parameter $(c_1,c_2)=(1,1-10^{-6})$ in Area2 of Fig.~\ref{fig:dual_arm_StaChart}.
    The green line represents the perturbed parameter $(c_1,c_2)=(1,1-10^{-6})$ in Area1 of Fig.~\ref{fig:dual_arm_StaChart}.
    The blue line represents the parameter $(c_1,c_2)=(1,1)$ point A of Fig.~\ref{fig:dual_arm_StaChart}}
    \label{fig:Bodeplot}
\end{figure}

As a result, the frequency response of the dual arm-locking system is given by
\begin{align}
C(s) &=G_L(s)G_c(s) \notag \\
& = g\left[\frac{s^2}{(s+p_{11})^2}\right]\times \notag \\
&\left[\frac{(s+z_{21})(s+z_{22})(s+z_{23})(s+z_{24})}{(s+p_{21})(s+p_{22})(s+p_{23})(s+z_{24})}\right] \tag{11}
\end{align}
The values of the zeros, poles and gain for each stage are summarized Table.~\ref{tab:dual_arm_controller}. As illustrated in Fig.~\ref{fig:Bodeplot}, the Bode plot of the dual arm-locking systems $C(s)P_d(s)$ under parameter perturbations confirm that the designed controller $C(s)$ ensures system stability for perturbations within $\pm10^{-6}$. Furthermore, the Bode plot indicates that the gain magnitude is virtually constant over the 0.01 mHz-1 Hz frequency range, a property that ensures effective noise suppression.
\begin{table}[htb]
    \centering
    \caption{Dual arm-locking controller parameters}
    \label{tab:dual_arm_controller}
    \begin{tabular}{lll}
    \hline
    Zeros(rad/s) & Poles(rad/s)  &Gain  \\
    \hline
    & $p_{11}=2\pi\times 10^{-6}$ & g=5000 \\
    $z_{21}=2\pi\times 10^{-8}$ & $p_{21}=2\pi\times 10^{-6}$ &  \\
    $z_{22}=2\pi\times 5\times 10^{-6}$ & $p_{22}=2\pi\times 5\times 10^{-5}$ &  \\
    $z_{23}=2\pi\times 5\times 10^{-6}$ & $p_{23}=2\pi\times 4\times 10^{-4}$ &  \\
    $z_{24}=2\pi\times 5\times 10^{-6}$ & $p_{24}=2\pi\times 5\times 10^{-5}$ &  \\
    \hline
    \end{tabular}
\end{table}
\begin{figure*}[t]
    \centering
    \begin{subfigure}{0.9\textwidth}
        \includegraphics[width=\linewidth]{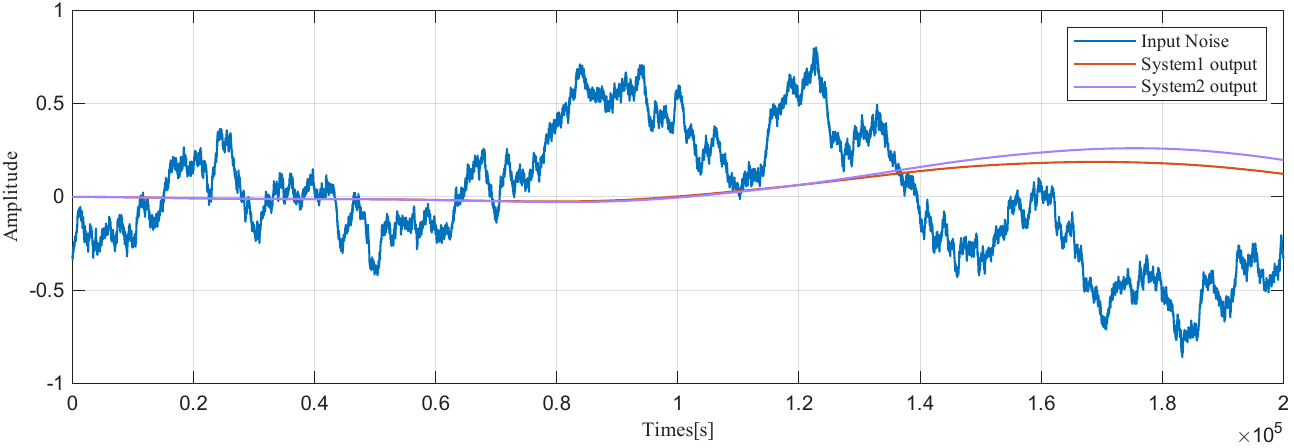}
        \caption{}
        \label{fig:simu_a}
    \end{subfigure}
    \begin{subfigure}{0.45\textwidth}
        \includegraphics[width=\linewidth]{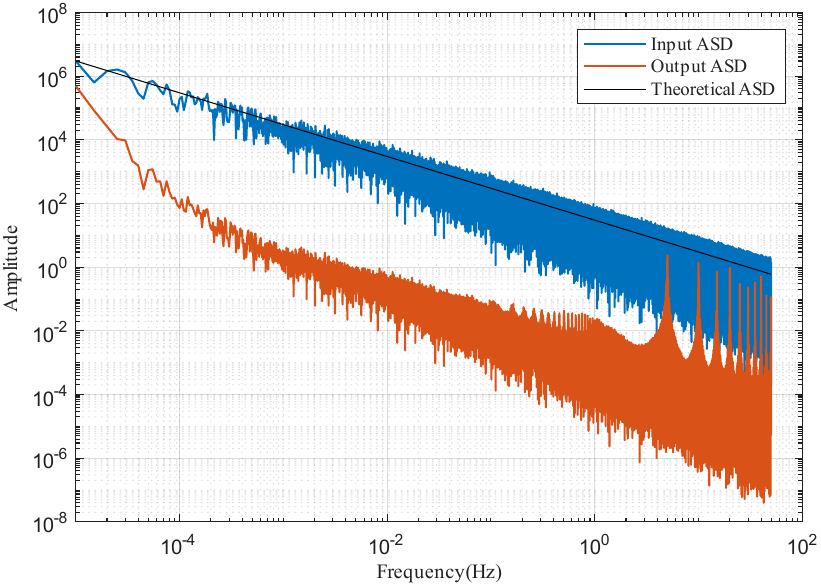}
        \caption{}
        \label{fig:simu_b}
    \end{subfigure}
    \begin{subfigure}{0.45\textwidth}
        \includegraphics[width=\linewidth]{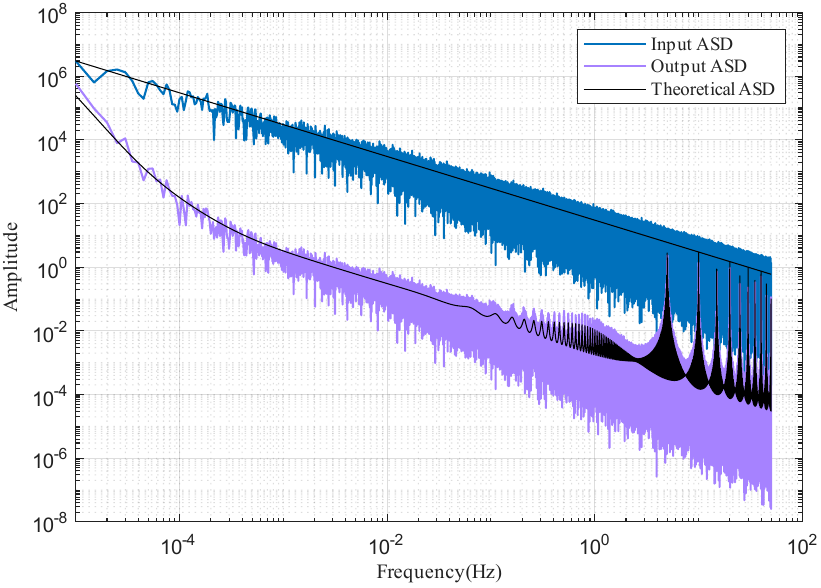}
        \caption{}
        \label{fig:simu_c}
    \end{subfigure}
    \captionsetup{
        justification=raggedright, 
        singlelinecheck=false       
    }
    \caption{Simulation results.
    (a) Time-domain simulation results of dual arm-locking system with controller $C(s)$ under parameter perturbations. Both System~1 and System~2 are driven by the same input (blue). System~1 is affected by a time-varying parameter (Fig.~\ref{fig:c1c2}) perturbation, whereas System~2 is subjected to a strong constant perturbation $(c_1,c_2)=(1,1-10^{-6})$. The orange and purple traces denote the outputs of System1 and System2, respectively.
    (b) Frequency-domain simulation results of system1 under time-varying parameter perturbation (Fig.~\ref{fig:c1c2}).
    (c) Frequency-domain simulation results of system 2 under parameter perturbation$(c_1,c_2)=(1,1)$.}
    \label{fig:SimulationResults}
\end{figure*}
\subsection{Time-domain Simulation}
A time-domain simulation of dual arm-locking system with the controller designed in Section~\ref{sec:controller_design} was conducted to verify both closed-loop stability and laser frequency noise suppression. The input laser frequency noise followed an ASD of $30/f\text{Hz}/\sqrt{\text{Hz}}$. The time -varying parameters perturbation $c_1,c_1$ were generated from its ASD using Eq.~(\ref{eq:c_ASD}) given in Appdendix~\ref{sec:appendix}, following ~\cite{dong2016comprehensive}. The discrete time simulation was performed at a sampling rate of $200$ Hz for a total duration 200000s.

Fig.~\ref{fig:SimulationResults} summarizes the system responses under different parameter perturbations. The system stays stable while preserving the achieved suppression under both time-varying and constant parameter perturbation. The proposed scheme reduces the laser frequency noise by approximately three to four orders of magnitude within $0.1$ mHz - $1$ Hz
The closed-loop outputs of System1(Fig.~\ref{fig:simu_b}) and System2(Fig.~\ref{fig:simu_c}) are nearly identical. This is because, under different parameter perturbations, the magnitude responses in the Bode plot of Fig.~\ref{fig:Bodeplot} are almost the same above $10^{-6}$Hz. As shown in the time-domain simulation results (Fig.~\ref{fig:simu_a}), the system outputs under different parameter perturbations exhibit noticeable deviations over long timescales. This behavior is consistent with the Bode plot analysis (Fig.~\ref{fig:Bodeplot}), in which the phase response shows small deviations at low frequencies. These low-frequency phase shifts accumulate over time, leading to the long-term differences observed in the time-domain responses. Furthermore, comparison with the theoretical prediction (black trace in Fig.~\ref{fig:simu_c}) indicates agreement between the Simulink simulation and the analytical model. Since System~1 is time-varying, its closed-loop theoretical response is not shown in Fig.~\ref{fig:simu_b}.
\section{Conclusion}
In this work, the stability of the arm-locking system for space-based gravitational-wave detectors has been analyzed. A parametric stability framework was established by combining the D-subdivision and Semi-Discretization methods. The analysis reveals that the arm-locking system operates near the stability boundary in the parameter space, making it highly sensitive to multiplicative perturbations. To mitigate this limitation, a robust controller architecture incorporating a high-pass filter was proposed. Although the formulation is similar to that used for suppressing Doppler-induced laser frequency pulling, our design requires careful parameter optimization to preserve overall closed loop stability.
\appendix
\section{Phase-locked parameters perturbation}
\label{sec:appendix}
\setcounter{equation}{0}
The laser frequency stabilization system of LISA is a complex engineering setup that 
involves multiple transformations from nonlinear to linear behavior. 
The PLL loop itself is nonlinear, and the coupling between clock noise and the phase demodulation process can lead to parameter perturbations. 
In addition, laser relative intensity noise (RIN), laser pointing noise (TTL), 
and residual laser phase noise in the PLL may cause the noise parameters $c_1$ and $c_2$ to deviate from their ideal values.
There gives a possible perturbation of the parameters $c_1$ and $c_2$ in the arm-locking system. 
The laser interference signals can be expressed as:
\begin{equation}
    \label{eq:laser_interference}
    S(t) = A\cos(\omega_{beat}+\varphi_m(t))
\end{equation}
Here, $A$ denotes the amplitude of the interference signal, $\omega_{\text{beat}}$ is the beat frequency, and $\varphi_m(t)$ represents the measurement phase. In the demodulation process, two orthogonal channels (I/Q) are typically used, and their low-pass filtered outputs are given by  
\begin{equation}
    \begin{cases}
        I_{\text{LPF}} = \dfrac{A(t)}{2}\,(1+\epsilon_I)\cos\bigl(\varphi(t)\bigr), \\
        Q_{\text{LPF}} = \dfrac{A(t)}{2}\,(1+\epsilon_Q)\sin\bigl(\varphi(t)\bigr).
    \end{cases}
\end{equation}
The measured phase is then calculated as
\begin{equation}
    \varphi_m(t) = \arctan\!\left(\dfrac{(1+\epsilon_Q)\sin(\varphi(t))}{(1+\epsilon_I)\cos(\varphi(t))}\right),
\end{equation}
where $\epsilon_Q, \epsilon_I \ll 1$. Using a Taylor series expansion, one obtains 
\[
1 + (\epsilon_Q - \epsilon_I) \approx 1 + \eta.
\]
Therefore, when there is a gain mismatch or imperfect quadrature between I and Q channels in the demodulation circuit, a multiplicative DC phase error will occur.

In this work, the considered phase measured model is given by
\begin{equation}
    \label{eq:A4}
    \varphi_m = (1+\eta)\varphi,
\end{equation}
where $\varphi_m$ denotes the laser phase measured by the phase measurement module or outputs of PPL module, $\eta$ represents the parameter perturbation associated with $c_1$ and $c_2$, and $\varphi$  denotes the true phase value. Similar multiplicative noise modeling can also be found in~\cite{schwarze2019picometer}. For LISA, the phase noise requirement typically modeled as additive, is expressed as
\begin{equation}
    \label{eq:c_ASD}
    \delta \varphi = 3.6\times10^{-5}\cdot\sqrt{1+\left(\frac{2.8\,\text{mHz}}{f}\right)^4}
    \quad \text{rad}/\sqrt{\text{Hz}}.
\end{equation}
For conservatism, a more cautious assumption is adopted in this work for the multiplicative noise term $\eta$. Specifically, the intensity of the additive noise is assumed to be comparable to that of the multiplicative noise; that is, the amplitude spectral density (ASD) of the additive noise  is used as an estimate for the ASD of $\eta$.

According to the model above, let $|\Phi(f)|^2 $ denote the PSD of $\varphi$, and $S_{\eta}$  denote the PSD of $\eta$. Then, the PSD of the measurement error $\delta\varphi = \varphi_m-\varphi$ can be expressed as $|\Phi(f)|^2 * S_{\eta}(f)$. The symbol * here represents convolution. It follows that $S_{\eta}$  typically needs to be one or several orders of magnitude smaller than $\delta\varphi$ to yield a measurement error comparable to that of the additive noise model. 
\begin{figure}[htp]
    \centering
    \includegraphics[width=0.98\linewidth]{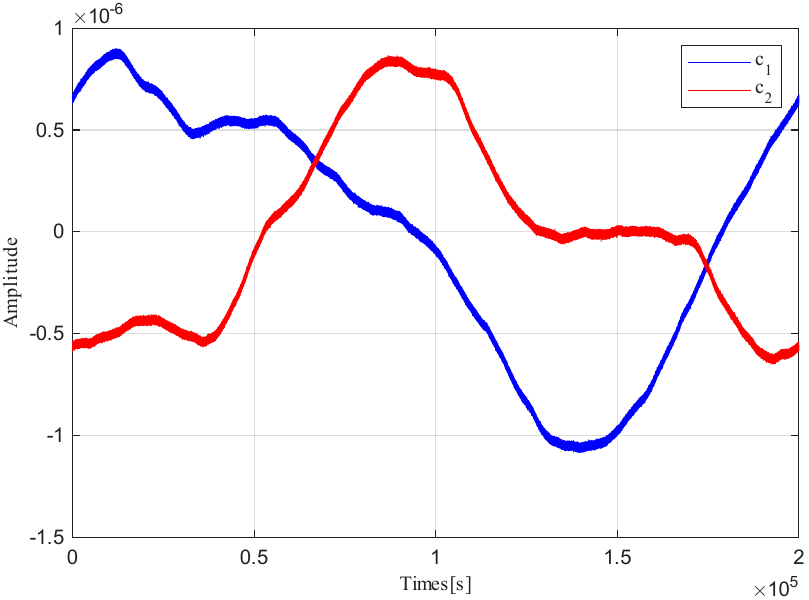}
    \caption{Time-varying parameter perturbations $c_1$ and $c_2$, generated as time-domain sequences from their amplitude spectral densities (ASD) according to Eq.~\ref{eq:c_ASD}.}
    \label{fig:c1c2}
\end{figure}
\bibliographystyle{apsrev4-2}
\bibliography{ref}
\end{document}